%% file: main.tex
\PassOptionsToPackage{table}{xcolor}
\documentclass{article} 
\usepackage{iclr2027_conference,times}

\usepackage{hyperref}
\usepackage{url}

\usepackage{graphicx}
\usepackage{booktabs}
\usepackage{amsfonts}
\usepackage{amsmath}
\usepackage{amssymb}
\usepackage{nicefrac}
\usepackage{microtype}
\usepackage{multirow}
\usepackage{xcolor}
\usepackage{subcaption}
\usepackage{wrapfig}
\usepackage{algorithm}
\usepackage{algpseudocode}
\usepackage{tikz}
\usetikzlibrary{arrows.meta}
\usepackage{framed}

\graphicspath{{figures/}}

\usepackage{longtable}
\usepackage{array}
\usepackage{calc}
\usepackage{adjustbox}
\usepackage{float}
\usepackage{needspace}
\usepackage[section,above,below]{placeins}
\usetikzlibrary{positioning,fit,backgrounds}

\hypersetup{unicode=true,colorlinks=true,linkcolor=blue!45!black,citecolor=blue!45!black,urlcolor=blue!45!black}

\definecolor{paperblue}{HTML}{345F80}

\newcommand{\resultcell}[2]{\makebox[3.6em][r]{#1}\makebox[.6em][l]{\textsuperscript{#2}}}

\title{Rethinking Latent Reasoning for\\Generative Recommendation:\\Beyond Teacher-Generated CoT and\\Recurrent Hidden-State Feedback}
\author{\begin{tabular}{@{}c@{}}
  Yiwen Chen$^{1,2}$ \hspace{0.5em}
  Fuwei Zhang$^{1}$ \hspace{0.5em}
  Zehao Chen$^{1}$ \hspace{0.5em}
  Deqing Wang$^{1}$ \hspace{0.5em}
  Hehan Li$^{2}$ \hspace{0.5em}
  Peizhi Xu$^{2}$ \\[4pt]
  Hanmeng Liu$^{2}$ \hspace{0.5em}
  Shuanglong Li$^{2}$ \hspace{0.5em}
  Xin Pei$^{2}$ \hspace{0.5em}
  Fuzhen Zhuang$^{1,\dagger}$ \hspace{0.5em}
  Zhao Zhang$^{1}$ \\[6pt]
  $^1$School of Artificial Intelligence, Beihang University, Beijing, China \\
  $^2$Baidu, Beijing, China \\
  \texttt{zhuangfuzhen@buaa.edu.cn}
\end{tabular}}
\iclrfinalcopy

\hypersetup{pdftitle={Rethinking Latent Reasoning for Generative Recommendation: Beyond Teacher-Generated CoT and Recurrent Hidden-State Feedback},pdfauthor={Yiwen Chen et al.}}
\begin{document}
\raggedbottom
\maketitle
\begin{abstract}
\input{sections/abstract}
\end{abstract}
\input{sections/01_introduction}
\input{sections/02_related_work}
\input{sections/03_cot_analysis}

\input{sections/04_method}
\input{sections/05_experiments}
\input{sections/08_conclusion}
\input{sections/07_limitations}
\bibliographystyle{iclr2027_conference}
\bibliography{references_lasar,refs_iclr,refs_new}
\clearpage
\appendix
\input{sections/appendix}
\end{document}

%% file: sections/abstract.tex
Latent reasoning in generative recommendation faces two practical barriers: constructing teacher-generated chain-of-thought (CoT) supervision is costly, and recurrent hidden-state feedback requires a specialized execution loop. We revisit both the supervision source and the computation interface. A step-level diagnosis measures how recommendation CoT supports next-item prediction and trace coherence. In the analyzed traces, history analysis is the only step with positive mean contributions on both dimensions, while later preference and intent steps have negative mean answer contributions. Guided by this finding, we propose \textbf{PHiT} (Prefilled, History-Informed Thought Tokens), which learns intermediate states from semantic targets extracted directly from observed item histories. PHiT aggregates cached sparse item encodings to construct these targets and supervises the contextual states of distinct vocabulary thought tokens. All thought tokens are supplied together as ordinary input IDs during a single causal prefill, removing the need for recurrent final-state feedback. With supervised fine-tuning alone, PHiT outperforms the compared baselines across four recommendation benchmarks. Controlled ablations support the value of input-matched semantic supervision, while cost measurements show fast target construction with training time comparable to the recommendation-only thought-token baseline.

%% file: sections/01_introduction.tex
\section{Introduction}
\label{sec:intro}

Generative recommendation formulates next-item prediction as autoregressive sequence generation, mapping a user's interaction history to the target item's semantic identifier (SID)~\citep{34_TIGER,35_IndexItemIDs,36_LCRec}. Reasoning-enhanced recommenders introduce intermediate computation to infer user preferences and purchase intent before predicting the next item~\citep{41_GREAM,42_RecSAVER,43_Reason4Rec,45_R2EC}. Latent reasoning methods internalize this intermediate process within hidden representations rather than decoding explicit text tokens; LASAR, for instance, aligns recurrent latent states with teacher Chain-of-Thought (CoT) representations and applies reinforcement learning (RL) to determine the reasoning depth~\citep{lasar2026}.

Avoiding rationale generation at test time leaves two practical costs: constructing training supervision and executing recurrent latent states. First, latent alignment remains tied to explicit teacher supervision: a large language model must still generate a reasoning trace for every training instance before the student can compress that trace into hidden states~\citep{08_Coconut,60_CODI,lasar2026}. Second, Coconut-style recurrent execution feeds a final-layer hidden state back as the next step's input embedding~\citep{08_Coconut}, requiring a feedback loop beyond the ordinary token-input prefill and decoding path used by serving engines such as vLLM~\citep{vllm2023}. We therefore examine the supervision source and the inference procedure (Figure~\ref{fig:motivation}).

\input{figures/motivation}

Which parts of a recommendation CoT trace are useful for supervision? A typical teacher trace interleaves factual evidence from observed interactions with extrapolated interpretations of user preferences and future intent, yet full-trace distillation treats all of this content as supervision. We develop a recommendation-specific diagnosis that measures each step's contribution to target-SID prediction and local trace coherence, then tests the resulting supervision choices through student retraining. Inspired by the step-utility perspective of SGP-CoT~\citep{sgpcot2026}, this analysis connects trace-level evidence to downstream recommendation quality. On Beauty, we observe that history analysis is the only reasoning step with positive mean contributions on both axes, whereas subsequent preference construction and intent inference yield negative mean answer contributions (Figure~\ref{fig:diagnosis}). Consistently, retaining only the history-analysis step on Instruments reduces supervision text by over 61\% while improving all six recommendation metrics over full-CoT supervision (Figure~\ref{fig:u1_grouped}). These findings motivate supervision grounded in observed historical evidence.

We introduce \textbf{PHiT} (Prefilled, History-Informed Thought Tokens) to supervise latent computation directly from observed item histories without teacher generation. PHiT encodes item descriptions once using a frozen sparse encoder, caches the resulting lexical representations, and aggregates them across each user's interaction history into a multi-hot semantic target. During training, a lightweight linear head predicts this historical target from intermediate thought-token states jointly with the next-item SID objective. Because item representations are cached and reused across all user histories, PHiT replaces per-example autoregressive teacher generation with offline item encoding and fast history aggregation, relying exclusively on observed historical items without accessing the target next-item label.

PHiT also changes how intermediate states are computed. We expand the model vocabulary with $N$ distinct thought tokens and insert them as a contiguous block between the interaction history and the target SID. Fixed thought-token IDs acquire user-specific contextual hidden states through causal self-attention over the preceding interaction history, retaining historical semantic evidence. Because both the sparse encoder and the auxiliary prediction head are discarded after training, inference simply appends the fixed thought-token IDs to the history prompt, computes their contextual states in a single causal prefill pass, and decodes the target SID through unmodified token and KV-cache interfaces.

Evaluated across Beauty, Instruments, Sports, and Steam using supervised fine-tuning (SFT) alone, PHiT achieves the best reported results on all 24 main-table metrics, with 22 statistically significant improvements over the strongest baselines. Controlled five-seed ablations on Beauty verify that supervising thought tokens with input-matched historical targets yields consistent gains beyond adding extra token positions or an unaligned auxiliary objective, while the complete target-construction pipeline processes 131,413 training examples in approximately 113 seconds on one GPU. Our contributions are:
\begin{itemize}
  \item \textbf{Diagnosing recommendation CoT supervision.} To our knowledge, we present the first step-level CoT supervision diagnosis for generative recommendation that combines target-SID likelihood, local trace coherence, and downstream student retraining. This analysis identifies historical evidence as an effective supervision source and motivates direct target construction from observed items.
  \item \textbf{Learning thought tokens from history.} We propose PHiT, which constructs reusable semantic targets from observed items to supervise vocabulary-based thought tokens, supporting teacher-free supervision and standard causal-prefill execution.
  \item \textbf{Validating recommendation quality and efficiency.} Through experiments across four benchmarks and five-seed controlled ablations, we demonstrate state-of-the-art recommendation accuracy under SFT alone while quantifying the substantial reduction in supervision construction cost.
\end{itemize}

%% file: figures/motivation.tex
\begin{figure}[!t]
\centering
\includegraphics[width=\linewidth]{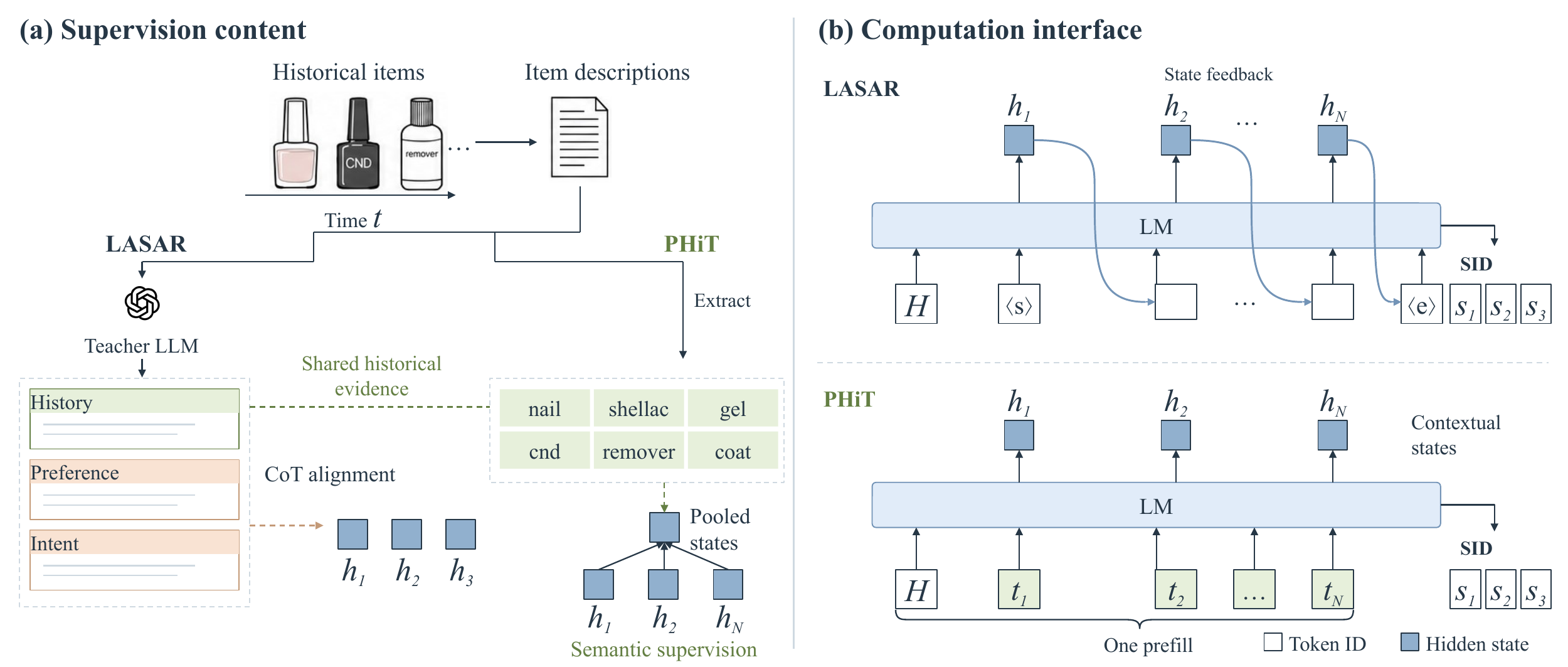}
\caption{\textbf{Rethinking the supervision source and computation interface.} (a) LASAR transfers a teacher-generated trace into latent states. We instead extract historical semantic targets directly from item text; the historical target example is detailed in Appendix~\ref{app:historical_example}. (b) LASAR feeds final-layer states into subsequent model evaluations. Our fixed token IDs are supplied together in prefill and develop history-dependent states before SID decoding. Here $H$ abbreviates the history-conditioned input $x_u$. Figure~\ref{fig:diagnosis} gives the step-level evidence.}
\label{fig:motivation}
\end{figure}

%% file: sections/02_related_work.tex
\section{Related Work}
\label{sec:related}

\paragraph{Generative and Reasoning-Aware Recommendation}

Sequential recommendation uses recurrent and self-attentive sequence encoders~\citep{28_SASRec,29_BERT4Rec,30_GRU4Rec} and language models that cast item prediction as conditional text generation~\citep{31_P5,32_M6Rec,33_TALLRec}. Within generative recommendation, discrete semantic indexing and collaborative-semantic alignment map items into hierarchical token sequences (SIDs) suitable for autoregressive decoding~\citep{34_TIGER,35_IndexItemIDs,36_LCRec,40_EAGER}, complemented by streamlined training and inference architectures~\citep{37_MiniOneRec,38_E4SRec}. Building on these generative backbones, reasoning-aware recommenders incorporate explicit rationale generation, preference modeling, and reinforcement learning (RL) to refine next-item predictions~\citep{41_GREAM,42_RecSAVER,43_Reason4Rec,44_RecR1,45_R2EC}, while latent and slow-thinking models allocate extra hidden computation prior to scoring or decoding~\citep{46_ReaRec,47_LARES,48_STREAMRec}. Among the most closely related generative methods, LatentR$^3$ avoids explicit CoT annotations by combining SFT with RL~\citep{51_LatentR3}, S$^2$GR interleaves thinking tokens with SID generation under codebook-cluster supervision~\citep{52_S2GR}, and LASAR aligns recurrent latent states with teacher CoT representations before optimizing reasoning depth via RL~\citep{lasar2026}. In contrast, PHiT eliminates both teacher-generated traces and RL post-training, supervising fixed input thought tokens with history-derived semantic targets through SFT alone.

\paragraph{From Explicit Reasoning to Latent Computation}

Knowledge distillation and progressive internalization compress explicit verbal reasoning into hidden states~\citep{09_ImplicitCoT,10_FromExplicitCoTtoImplicitCoT,60_CODI}. Continuous latent reasoning methods such as Coconut feed final-layer hidden states back as subsequent input embeddings~\citep{08_Coconut}, inspiring extensions based on compressed traces, distilled multi-step computation, and continuous thinking representations~\citep{13_CCoT,14_System2to1,17_Heima}, as well as soft token projections, latent-state optimization, and adaptive iteration depth~\citep{18_SoftCoT,19_LaTRO,20_ThinkAtHard}. Concurrently, architectural studies show that intermediate computation need not mirror natural-language traces: pause and filler tokens allocate extra forward positions~\citep{11_PauseTokens,12_ThinkDotByDot}, looped transformers and expressivity analyses characterize the representational gains of intermediate depth~\citep{07_ExpressivePower,15_LoopedTransformers,16_Huginn}, and discrete latent tokens learned from reasoning traces can be interleaved with text~\citep{21_TokenAssorted}. PHiT uses discrete vocabulary positions for intermediate computation and supervises their hidden states with semantic targets extracted directly from observed interaction histories.

\paragraph{Reasoning Utility and Efficient Supervision}

The benefits of CoT prompting and test-time search depend on step quality and compute allocation~\citep{01_CoT,02_ZeroShot,03_SelfConsistency,04_ToT,24_TestTimeCompute}. Recent work therefore examines whether every generated reasoning step contributes positively to the final prediction. Specifically, SGP-CoT quantifies the answer and coherence contributions of individual reasoning steps to prune redundant segments from generated traces~\citep{sgpcot2026}. Inspired by step-utility analysis, our recommendation-specific protocol evaluates target-SID prediction and local trace coherence, and connects these measurements to student retraining. This analysis guides the selection of supervision content and leads to history-derived targets that bypass teacher-trace generation during target construction.

\paragraph{Sparse Evidence and Token-Based Execution}

Sparse lexical encoders such as SPLADE project text into high-dimensional vocabulary weight vectors that capture salient semantic terms~\citep{splade2021,spladev2_2021}. UEmbed extends sparse lexical representations to a decoder-only multimodal model, using learnable special tokens assigned to disjoint vocabulary subsets to produce sparse and dense embeddings in a single causal forward pass~\citep{song2026uembed}. PHiT instead uses a pooled thought-token representation to predict history-derived semantic targets for next-item recommendation. PHiT uses these lexical weights offline to construct reusable, history-grounded auxiliary targets from item metadata. At inference time, modern serving engines such as vLLM rely on standard token-level causal prefill and paged KV-cache management for throughput efficiency~\citep{vllm2023}. Because PHiT confines sparse encoding and keyword prediction to offline target construction and training, its test-time execution reduces to supplying fixed thought-token IDs within the recommendation model's native causal prefill path.

%% file: sections/03_cot_analysis.tex
\section{Diagnosing the Value of Recommendation CoT}
\label{sec:analysis}

\subsection{Diagnosing the Supporting Content}

We first examine which steps in a teacher trace support predicting the target item SID and which support continuing the verbal narrative (Figure~\ref{fig:diagnosis}).

\input{figures/diagnosis}

\paragraph{Step-level measurements.}

Given a user's interaction history $H_u=(i_1,\ldots,i_T)$, the recommendation model predicts the next item's SID $y_u=(s_1,\ldots,s_L)$ from the history-conditioned prompt $x_u$. In explicit-CoT supervision, the teacher provides a structured reasoning trace $C_u=(U_1,U_2,U_3)$ segmented by explicit \texttt{Step N} markers into three multi-sentence functional units: \textbf{history analysis} ($U_1$), \textbf{preference construction} ($U_2$), and \textbf{intent inference} ($U_3$). This three-step template follows the explicit teacher-CoT protocol established by GREAM and LASAR, in which an LLM teacher analyzes the observed history, constructs plausible user preferences, and infers purchase intent before the target SID is generated~\citep{41_GREAM,lasar2026}. Both Explicit CoT$_{\text{GREAM}}$ and LASAR in Table~\ref{tab:main_results} use this form of supervision.

Our diagnosis evaluates two forms of support in recommendation traces: support for the target item SID and support for the next reasoning unit. Inspired by the step-deletion perspective of SGP-CoT~\citep{sgpcot2026}, we instantiate answer contribution with length-normalized SID likelihood and coherence contribution with next-unit likelihood. We then evaluate the supervision design through student retraining. For a target SID $y$ and context $z$, we define
\begin{equation}
\ell_A(y\mid z)=\frac{1}{L}\sum_{j=1}^{L}\log p_\theta(s_j\mid z,s_{<j}).
\end{equation}
We measure the answer contribution of $U_i$ as the difference in target-SID log-likelihood between the full trace and the trace with $U_i$ removed:
\begin{equation}
\Delta A_i=\ell_A(y\mid x,C)-\ell_A(y\mid x,C\setminus U_i).
\label{eq:answer_contribution}
\end{equation}
For any non-terminal unit ($i < 3$), we similarly measure its coherence contribution as the change in length-normalized log-likelihood of the immediately following reasoning unit $U_{i+1}$ when $U_i$ is included versus omitted from the prefix:
\begin{equation}
\Delta C_i=\ell(U_{i+1}\mid x,U_{1:i})-\ell(U_{i+1}\mid x,U_{1:i-1}),
\label{eq:coherence_contribution}
\end{equation}
where $\ell$ is normalized by the token length of $U_{i+1}$, and $\Delta C_3$ is undefined because the terminal unit $U_3$ has no successor. Positive values of $\Delta A_i$ and $\Delta C_i$ indicate that $U_i$ facilitates predicting the recommendation answer and continuing the subsequent reasoning step, respectively. We evaluate both metrics using the recommendation-trained Stage~1 student model with a no-CoT reference check to separate genuine support for next-item prediction from internal narrative continuity.

\paragraph{Historical evidence supports both predictions.}

On the Beauty training set, history analysis ($U_1$) is the only step that achieves positive mean contributions on both dimensions, yielding $\Delta A_1=+0.0183$ and $\Delta C_1=+0.0344$ (Figure~\ref{fig:diagnosis}; full means and medians in Appendix~\ref{app:diagnosis}). By contrast, preference construction ($U_2$) exhibits negative mean contributions to both answer prediction ($\Delta A_2=-0.0627$) and trace coherence ($\Delta C_2=-0.0766$), and intent inference ($U_3$) similarly reduces target-answer likelihood ($\Delta A_3=-0.0307$). These measurements motivate supervising intermediate states with observed historical evidence.

\subsection{Does U1-Only Supervision Improve Recommendation on Instruments?}

We conduct SFT-only experiments within the LASAR framework on Amazon Instruments, comparing full-CoT supervision with supervision using only the history-analysis step ($U_1$). As shown in Figure~\ref{fig:u1_grouped}, U1-only supervision reduces the average teacher trace length by over 61\% while improving all six recommendation metrics over full-CoT supervision (settings in Appendix~\ref{app:u1_observation}). This comparison supports retaining historical evidence as intermediate supervision.

\input{figures/alternatives/u1_grouped}

\subsection{From Historical Content to Direct Supervision}

The diagnosis identifies grounded historical evidence as useful supervisory content, and the downstream comparison shows that retaining this content can improve recommendation with shorter teacher traces. This motivates a further step: construct semantic targets directly from observed item descriptions, bypassing teacher generation altogether. Section~\ref{sec:method} develops this approach, and the fixed-$N$ ablations in Section~\ref{sec:experiments} compare history-derived targets with alternative supervision sources under a shared thought-token interface.

%% file: figures/diagnosis.tex
\begin{figure}[!htb]
\centering
\includegraphics[width=\linewidth]{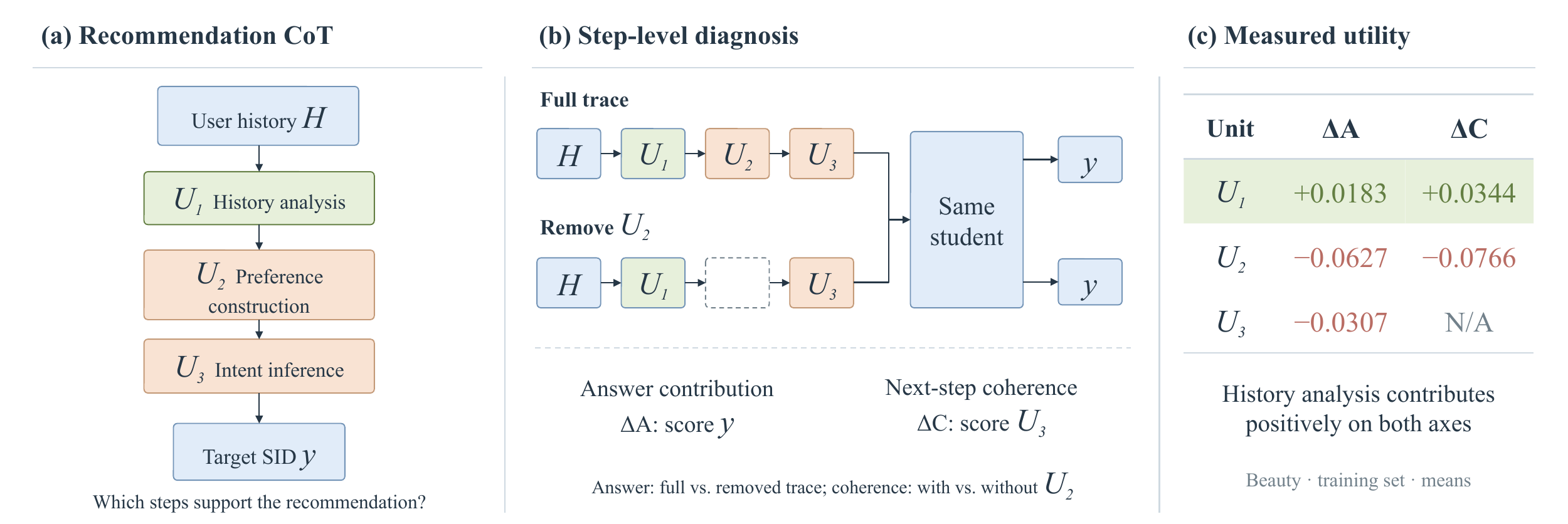}
\caption{\textbf{Diagnosing the contribution of each CoT step.} (a) History analysis ($U_1$), preference construction ($U_2$), and intent inference ($U_3$). (b) The shown traces compare target-SID likelihood before and after removing $U_2$. Coherence rescoring predicts $U_3$ from the prefix with versus without $U_2$, using the same Stage~1 student. (c) Mean contributions over CoT traces on the Beauty training set; $\Delta C_3$ is undefined because $U_3$ has no successor.}
\label{fig:diagnosis}
\end{figure}

%% file: figures/alternatives/u1_grouped.tex
\begin{figure}[!htb]
\centering
\includegraphics[width=\linewidth]{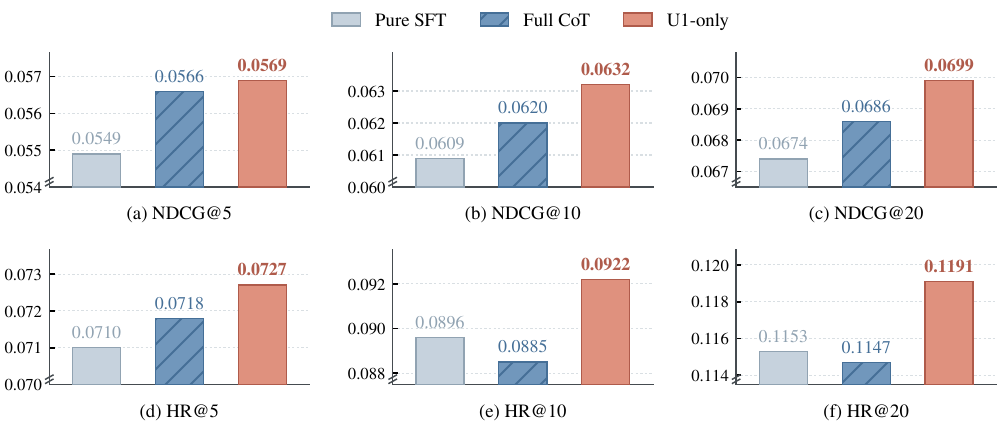}
\caption{\textbf{U1-only improves all six metrics with over 61\% less supervision text on Instruments.} The comparison uses the LASAR framework with SFT alone. Labels report absolute scores; double slashes mark truncated axes. U1-only retains the history-analysis step.}
\label{fig:u1_grouped}
\end{figure}

%% file: sections/04_method.tex
\section{PHiT}
\label{sec:method}

Guided by our diagnostic findings, \textbf{PHiT} (Prefilled, History-Informed Thought Tokens) supervises intermediate latent states using semantic targets extracted directly from observed item histories, while executing those states through discrete vocabulary tokens within a standard causal prefill pass. Figure~\ref{fig:overview} illustrates the complete workflow across offline target construction, joint supervised training, and standard-prefill inference. Following established generative recommendation protocols~\citep{34_TIGER,61_HandbookSID}, we instantiate PHiT on a Qwen3 backbone~\citep{qwen3} for autoregressive SID prediction.

\input{figures/overview}

\subsection{Constructing Targets from Historical Evidence}

We supervise thought-token states through multi-label keyword prediction using cached sparse item encodings. For each item $i$ in the catalog, we concatenate its title, brand, categories, and text description into a unified document $d_i$ and pass it through a frozen SPLADE encoder to obtain vocabulary-level lexical weights $a_i(w)$~\citep{splade2021,spladev2_2021}. We then merge subword WordPiece fragments into whole words, normalize casing, strip formatting artifacts, and filter the output to retain only words that appear verbatim in $d_i$. We compute each item representation once and cache it offline.

Given a user's observed interaction history $H_u$, we aggregate the cached lexical weights of its constituent items and select the top-$K$ highest-scoring words as the user's historical semantic summary:
\begin{equation}
q_u(w)=\sum_{i\in H_u}a_i(w),\qquad K_u=\operatorname{TopK}_w q_u(w).
\label{eq:historical_target}
\end{equation}
We represent $K_u$ as a multi-hot target vector $z_u \in \{0,1\}^{|\mathcal V_k|}$ over the training keyword vocabulary $\mathcal V_k$, setting $z_{u,w}=1$ if $w\in K_u$ and $0$ otherwise, with $K=16$ throughout our experiments. Constructing $z_u$ from the observed historical items in $H_u$ replaces $S$ conditional text generations with $M$ reusable item encodings and lightweight summation over user histories.

\subsection{Learning History-Supervised Thought Tokens}

To allocate intermediate computation while preserving a standard token-level interface, we expand the model vocabulary with $N$ distinct thought tokens $t_1,\ldots,t_N$ equipped with trainable input embeddings, and insert them between the history prompt $x_u$ and the target SID $y_u$ delimited by boundary markers:
\begin{equation}
[x_u,\langle\mathrm{start}\rangle,t_1,\ldots,t_N,\langle\mathrm{end}\rangle,y_u].
\label{eq:slot_sequence}
\end{equation}
Here $N$ is treated as a hyperparameter selected on validation data (see Figure~\ref{fig:b0_slot_scan} for a sensitivity study on Beauty). Although the token IDs $t_{1:N}$ are identical across all examples, causal self-attention over the preceding prompt $x_u$ produces user-specific final-layer hidden states $h_{u,1},\ldots,h_{u,N}$. During training, we mean-pool these thought-token states and apply a lightweight linear classification head to predict the historical semantic target $z_u$:
\begin{equation}
\bar h_u=\frac{1}{N}\sum_{n=1}^{N}h_{u,n},\qquad
\hat z_u=\sigma(W_k\bar h_u+b_k).
\end{equation}
Because the positive keyword set $\mathcal P_u = K_u$ is much smaller than the negative vocabulary set $\mathcal N_u = \mathcal V_k \setminus K_u$, we supervise $\hat z_u$ using a class-balanced binary cross-entropy loss so that negative terms do not dominate the gradient because of their greater number:
\begin{equation}
\mathcal L_{\mathrm{hist}}(u)=
-\frac{1}{|\mathcal P_u|}\sum_{w\in\mathcal P_u}\log\hat z_{u,w}
-\frac{1}{|\mathcal N_u|}\sum_{w\in\mathcal N_u}\log(1-\hat z_{u,w}).
\end{equation}
We optimize the model end to end via supervised fine-tuning by combining this auxiliary historical objective with standard autoregressive SID prediction:
\begin{align}
\mathcal L_{\mathrm{rec}}(u)&=-\sum_{j=1}^{L}\log p_\theta(s_j\mid x_u,t_{1:N},s_{<j}),\\
\mathcal L&=\mathbb E_u[\mathcal L_{\mathrm{rec}}(u)+\lambda\mathcal L_{\mathrm{hist}}(u)],
\end{align}
where the conditioning context in $\mathcal L_{\mathrm{rec}}(u)$ includes the boundary markers in Equation~\ref{eq:slot_sequence}, next-token language-modeling labels are masked on $x_u$ and the thought-token block, and $\lambda=1.0$. Through this joint objective, gradients from both historical keyword reconstruction and next-item SID decoding shape the intermediate thought-token states without requiring reinforcement learning.

\subsection{Executing the Thought Tokens in a Standard Prefill}

At inference time, both the frozen sparse encoder and the auxiliary keyword prediction head are discarded. For each user request, we append the fixed sequence of thought-token IDs $[\langle\mathrm{start}\rangle,t_1,\ldots,t_N,\langle\mathrm{end}\rangle]$ to the history prompt $x_u$, execute a single standard causal prefill over the combined sequence, and autoregressively decode the target SID $y_u$.

\paragraph{Fixed identities, input-dependent states.}
Within each transformer layer $\ell$, the hidden state of thought token $n$ is updated via causal self-attention over the history prompt, boundary tokens, and preceding thought tokens from layer $\ell-1$:
\begin{equation}
h_{u,n}^{(\ell)}=F_\theta^{(\ell)}\!\left(h_{u,n}^{(\ell-1)},\,H_{u,\leq n}^{(\ell-1)}\right).
\end{equation}
The same token IDs $t_{1:N}$ therefore produce contextual states $h_{u,1:N}^{(\ell)}$ that depend on the user's prompt $x_u$. During training, $\mathcal L_{\mathrm{hist}}$ encourages these contextual states to distill salient historical item semantics, while $\mathcal L_{\mathrm{rec}}$ organizes that information to directly support subsequent SID decoding.

\paragraph{Execution and construction complexity.}
Whereas Coconut-style recurrent latent reasoning~\citep{08_Coconut} must extract a final-layer hidden state and feed it back as the next step's input embedding across sequential latent updates, PHiT supplies all $N$ thought tokens simultaneously as part of the input prompt (Figure~\ref{fig:overview}). Adding $N$ thought tokens therefore increases the prefill sequence length and KV-cache footprint by $N$ positions, but requires zero autoregressive intermediate token generation and no custom hidden-state feedback loop, preserving standard token-input and KV-cache interfaces. Offline, for a catalog of $M$ items and $S$ training interactions, PHiT requires only $M$ sparse item encodings and $S$ fast vector aggregations, compared with $S$ full LLM generation calls for teacher-trace distillation (quantified in Section~\ref{sec:cost} and summarized in Algorithm~\ref{alg:pipeline}).

%% file: figures/overview.tex
\begin{figure}[!htb]
\centering
\includegraphics[width=\linewidth]{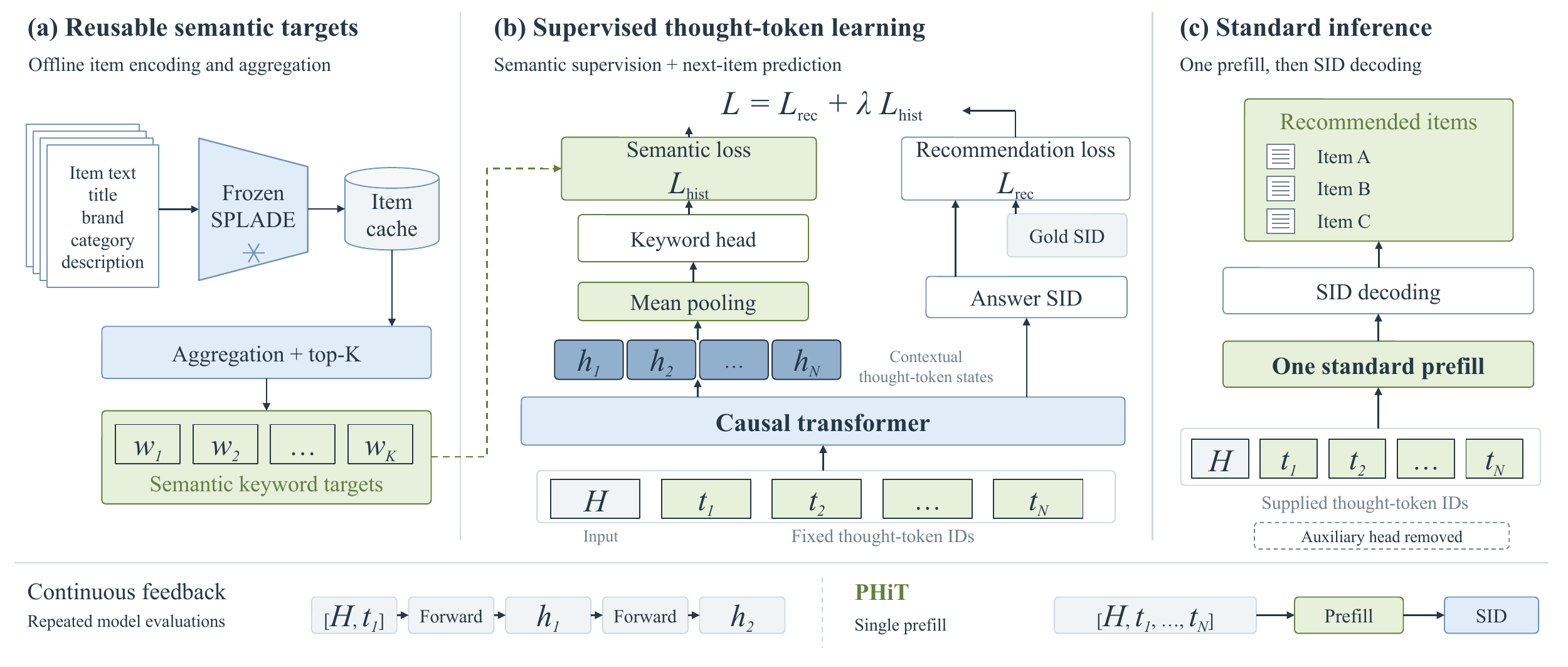}
\caption{\textbf{PHiT: semantic target construction, training, and inference.} (a) Cached item encodings supply keyword targets. (b) A training-only head predicts these targets from mean-pooled thought-token states, jointly with next-item prediction. (c) Supplied thought-token IDs use standard prefill and SID decoding. Here $H$ abbreviates $x_u$; user subscripts and boundary tokens are omitted.}
\label{fig:overview}
\end{figure}

%% file: sections/05_experiments.tex
\section{Experiments}
\label{sec:experiments}

We evaluate recommendation accuracy on four benchmarks, the effect of input-matched supervision through controlled Beauty ablations, and supervision construction and Stage~2 training time.

\subsection{Experimental Setup}

\paragraph{Datasets and baselines.}
We conduct next-item SID prediction experiments on four benchmarks: Amazon Beauty, Instruments, and Sports using the preprocessed splits released by GREAM~\citep{41_GREAM}, and Steam following the experimental setup of LASAR~\citep{lasar2026}. Table~\ref{tab:datasets} summarizes the dataset statistics (numbers of users, items, interactions, sparsity, and held-out test instances) from the shared LASAR benchmark inventory. We compare PHiT against seven representative baselines spanning traditional sequential recommenders (SASRec~\citep{28_SASRec} and GRU4Rec~\citep{30_GRU4Rec}), non-reasoning generative recommenders (LC-Rec~\citep{36_LCRec} and MiniOneRec~\citep{37_MiniOneRec}), latent sequential reasoning (ReaRec~\citep{46_ReaRec}), explicit CoT supervision (Explicit CoT$_{\text{GREAM}}$~\citep{41_GREAM}), and the full SFT-plus-RL latent reasoning framework LASAR~\citep{lasar2026}. All baseline scores in Table~\ref{tab:main_results} are cited directly from the LASAR benchmark, whereas PHiT is trained using SFT alone.

\input{tables/dataset_statistics}

\input{tables/table_02}

\paragraph{Evaluation protocol.}
Following the standard leave-one-out protocol in LASAR, the final interaction of each user sequence is held out for testing and a separate validation split is used for hyperparameter selection. All hyperparameters for the main experiments are selected on the validation split and fixed prior to test evaluation. Models decode ranked candidate lists of item SIDs via beam search, and we report Normalized Discounted Cumulative Gain (NDCG@$k$) and Hit Rate (HR@$k$) at cutoffs $k \in \{5, 10, 20\}$ for the main four-dataset comparison, supplemented by a finer top-rank (@1) cutoff in our controlled Beauty ablations. Further dataset and metric specifications appear in Appendix~\ref{app:settings}.

\paragraph{Implementation details.}
We instantiate PHiT on Qwen3-0.6B, initializing all matched Beauty ablation variants from an identical Stage~1 recommendation checkpoint. Offline historical targets are extracted using frozen \texttt{naver/splade\_v2\_max}~\citep{splade2021,spladev2_2021} encodings with $K=16$ top keywords per user history. The thought-token count is tuned over $N\in\{1,2,4,8\}$ on validation data; on Beauty, the selected configuration uses $N=8$, historical loss weight $\lambda=1.0$, and decoding beam width 50. In our controlled ablations, the \emph{+ CoT alignment} variant supervises thought-token hidden states against teacher CoT representations using a mean-squared error (MSE) loss weighted by $0.1$, whereas LASAR in Table~\ref{tab:main_results} denotes the complete two-stage SFT-and-RL pipeline.

\subsection{Recommendation Performance}
\label{sec:main_results}

\paragraph{SFT alone delivers consistent gains across all four domains.}
As shown in Table~\ref{tab:main_results}, PHiT achieves the highest performance across all 24 dataset--metric combinations spanning Beauty, Instruments, Sports, and Steam, with 22 of these 24 improvements reaching statistical significance over the strongest respective baseline. PHiT surpasses both explicit CoT supervision (Explicit CoT$_{\text{GREAM}}$) and the recurrent SFT-plus-RL framework LASAR using supervised fine-tuning alone, showing stronger recommendation quality without teacher-trace generation or an RL stage.

\paragraph{Strong gains across ranking cutoffs.}
On Sports, PHiT improves HR@20 by $21.7\%$ and NDCG@20 by $12.2\%$ over LASAR, extending its advantage to deeper ranking cutoffs. On Steam, it improves HR@5 by $11.2\%$, strengthening top-ranked recommendations. These gains accompany improvements across all six metrics on both Beauty and Instruments.

\subsection{Effectiveness of Historical Supervision}

Table~\ref{tab:t4} separates component additions from supervision choices on Beauty. MiniOneRec (SFT) is the recommendation-only reference without thought tokens; adding eight thought tokens gives \emph{Base}. The lower block independently adds one auxiliary objective to Base: \emph{+ Shuffled targets} predicts historical targets permuted across users; \emph{+ CoT alignment} aligns states with teacher-trace representations; \emph{+ CoT keywords} predicts teacher-derived keywords; and \textbf{PHiT} uses semantic targets matched to the observed history. The five thought-token conditions share the same Stage~1 initialization.

\input{figures/alternatives/beauty_evidence}

\paragraph{Thought tokens and semantic supervision improve the reference.}
PHiT improves NDCG@1 by $20.6\%$ and NDCG@5 by $10.3\%$ over MiniOneRec (SFT), with gains of $8.6\%$ on NDCG@10 and $8.1\%$ and $6.7\%$ on HR@5 and HR@10. Table~\ref{tab:t4} shows the component progression: adding thought tokens improves the recommendation-only reference, and matched semantic supervision further improves all five metrics over Base. We next compare the supervision sources under the same thought-token configuration.

\paragraph{Input-matched historical targets drive the supervision gain.}
The controlled five-variant comparison fixes the thought-token count. PHiT outperforms shuffled-target supervision on NDCG@5 ($p=0.013$). These two conditions share the architecture, head, objective, and target coverage, identifying the contribution of matched semantic targets. Figure~\ref{fig:core_ablation} compares the five conditions; PHiT exceeds Base and shuffled targets in every matched run. Appendix~\ref{app:seed_consistency} shows the run-level scores and paired gains.

\paragraph{Direct historical targets outperform teacher-derived supervision.}
Across all five evaluation metrics in Table~\ref{tab:t4}, PHiT achieves the highest mean score, outperforming both continuous representation alignment (\emph{+ CoT alignment}) and discrete keyword supervision extracted from teacher traces (\emph{+ CoT keywords}). Relative to \emph{+ CoT alignment}, PHiT improves NDCG@1 by $5.0\%$ and achieves higher NDCG@5 in all five matched runs ($p=0.018$). Similarly, PHiT outperforms \emph{+ CoT keywords} across all cutoffs ($p\approx0.083$ on NDCG@5). These results support our diagnostic hypothesis: extracting targets directly from observed items provides more effective intermediate supervision than the compared teacher-derived targets.

\Needspace{13\baselineskip}
\noindent\textbf{Sensitivity to the number of thought tokens.}\par\smallskip
\noindent\begin{minipage}[c]{.64\linewidth}
Figure~\ref{fig:b0_slot_scan} examines Base on Beauty with $N\in\{1,2,4,8\}$ and recommendation supervision alone. Mean test NDCG@5 scores are 0.02271, 0.02263, 0.02218, and 0.02356, respectively. The test-set trend agrees with the validation-set trend, with $N=8$ highest on both splits.
\end{minipage}\hfill
\begin{minipage}[c]{.32\linewidth}
\centering
\includegraphics[width=\linewidth]{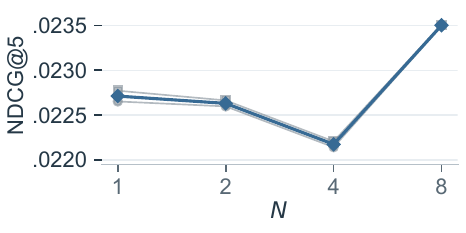}
\captionsetup{hypcap=false,skip=3pt}
\captionof{figure}{\textbf{Sensitivity to $N$ on Beauty.}}
\label{fig:b0_slot_scan}
\end{minipage}
\par\smallskip

\Needspace{24\baselineskip}
\subsection{Supervision and Training Efficiency}
\label{sec:cost}

Table~\ref{tab:cost_main} compares the offline supervision construction cost and Stage~2 training time of PHiT against teacher-CoT baselines on Beauty (hardware specifications and timing protocols in Appendix~\ref{app:cost_details}).

\input{figures/alternatives/cost}

\paragraph{Reusable item encodings accelerate supervision construction.}
Generating teacher CoT traces for all 131,413 Beauty training examples with Qwen3.6-35B-A3B~\citep{qwen36model} takes 145 minutes on eight A800 GPUs (66.20 ms per trace amortized). PHiT encodes each catalog item once and reuses its cached sparse weights across user histories, constructing all 131,413 training targets in 113 seconds on a single GPU (0.86 ms per target amortized).

\paragraph{Historical supervision preserves comparable Stage~2 training time.}
During Stage~2 fine-tuning on eight GPUs, Base and PHiT take 2,467 and 2,471 seconds, respectively, showing comparable training time while PHiT achieves higher recommendation accuracy. PHiT also achieves higher recommendation accuracy than \emph{+ CoT keywords} while reducing Stage~2 training time from 2,741 seconds to 2,471 seconds. At test time, neither the sparse encoder nor the keyword head is used; the additional execution work comes from the supplied thought-token positions.

%% file: tables/dataset_statistics.tex
\begin{table}[!htb]
\centering
\caption{Dataset statistics reported for the shared LASAR benchmark. Test examples correspond to held-out next-item predictions.}
\label{tab:datasets}
\small
\begin{tabular*}{\linewidth}{@{\extracolsep{\fill}}lrrrrr@{}}
\toprule
Dataset & Users & Items & Interactions & Sparsity & Test examples \\
\midrule
Beauty & 22,363 & 12,101 & 176,139 & 99.935\% & 22,363 \\
Instruments & 24,772 & 9,922 & 74,316 & 99.970\% & 24,772 \\
Sports & 35,598 & 18,357 & 106,794 & 99.984\% & 35,598 \\
Steam & 11,472 & 3,478 & 212,303 & 99.468\% & 11,472 \\
\bottomrule
\end{tabular*}
\end{table}

%% file: tables/table_02.tex
\begin{table}[!t]
\centering
\caption{Recommendation results on four datasets. PHiT uses SFT alone. Best values are \textbf{bold}; second-best values are \underline{underlined}.}
\label{tab:main_results}
\label{tab:t2}
\small
\setlength{\tabcolsep}{3pt}
\renewcommand{\arraystretch}{1.03}
\begin{tabular*}{\linewidth}{@{\extracolsep{\fill}}llrrrrrr@{}}
\toprule
 & & \multicolumn{3}{c}{NDCG $\uparrow$} & \multicolumn{3}{c}{HR $\uparrow$} \\
\cmidrule(lr){3-5}\cmidrule(l){6-8}
Dataset & Model & @5 & @10 & @20 & @5 & @10 & @20 \\
\midrule
\multirow{8}{*}{\rotatebox[origin=c]{90}{Sports}} & Explicit CoT$_{\text{GREAM}}$ & \resultcell{0.0089}{} & \resultcell{0.0118}{} & \resultcell{0.0153}{} & \resultcell{0.0138}{} & \resultcell{0.0228}{} & \resultcell{0.0370}{} \\
 & MiniOneRec & \resultcell{0.0099}{} & \resultcell{0.0126}{} & \resultcell{0.0152}{} & \resultcell{0.0155}{} & \resultcell{0.0237}{} & \resultcell{0.0339}{} \\
 & ReaRec & \resultcell{0.0086}{} & \resultcell{0.0112}{} & \resultcell{0.0143}{} & \resultcell{0.0151}{} & \resultcell{0.0233}{} & \resultcell{0.0355}{} \\
 & LC-Rec & \resultcell{0.0081}{} & \resultcell{0.0100}{} & \resultcell{0.0118}{} & \resultcell{0.0123}{} & \resultcell{0.0184}{} & \resultcell{0.0254}{} \\
 & GRU4Rec & \resultcell{0.0062}{} & \resultcell{0.0080}{} & \resultcell{0.0101}{} & \resultcell{0.0090}{} & \resultcell{0.0147}{} & \resultcell{0.0232}{} \\
 & SASRec & \resultcell{0.0060}{} & \resultcell{0.0074}{} & \resultcell{0.0094}{} & \resultcell{0.0089}{} & \resultcell{0.0132}{} & \resultcell{0.0212}{} \\
 & LASAR & \resultcell{\underline{0.0121}}{} & \resultcell{\underline{0.0152}}{} & \resultcell{\underline{0.0188}}{} & \resultcell{\underline{0.0185}}{} & \resultcell{\underline{0.0280}}{} & \resultcell{\underline{0.0425}}{} \\
\cmidrule(l){2-8}
\rowcolor{paperblue!6}
 & \textbf{PHiT (ours)} & \resultcell{\textbf{0.0122}}{} & \resultcell{\textbf{0.0161}}{*} & \resultcell{\textbf{0.0211}}{*} & \resultcell{\textbf{0.0198}}{*} & \resultcell{\textbf{0.0317}}{*} & \resultcell{\textbf{0.0517}}{*} \\
\midrule
\multirow{8}{*}{\rotatebox[origin=c]{90}{Instruments}} & Explicit CoT$_{\text{GREAM}}$ & \resultcell{0.0574}{} & \resultcell{0.0621}{} & \resultcell{0.0674}{} & \resultcell{0.0703}{} & \resultcell{0.0850}{} & \resultcell{0.1060}{} \\
 & MiniOneRec & \resultcell{0.0604}{} & \resultcell{0.0640}{} & \resultcell{0.0677}{} & \resultcell{0.0715}{} & \resultcell{0.0826}{} & \resultcell{0.0974}{} \\
 & ReaRec & \resultcell{0.0494}{} & \resultcell{0.0548}{} & \resultcell{0.0604}{} & \resultcell{0.0705}{} & \resultcell{0.0873}{} & \resultcell{0.1095}{} \\
 & LC-Rec & \resultcell{0.0533}{} & \resultcell{0.0561}{} & \resultcell{0.0587}{} & \resultcell{0.0616}{} & \resultcell{0.0701}{} & \resultcell{0.0803}{} \\
 & SASRec & \resultcell{0.0449}{} & \resultcell{0.0475}{} & \resultcell{0.0502}{} & \resultcell{0.0536}{} & \resultcell{0.0617}{} & \resultcell{0.0725}{} \\
 & GRU4Rec & \resultcell{0.0422}{} & \resultcell{0.0454}{} & \resultcell{0.0489}{} & \resultcell{0.0527}{} & \resultcell{0.0629}{} & \resultcell{0.0769}{} \\
 & LASAR & \resultcell{\underline{0.0612}}{} & \resultcell{\underline{0.0667}}{} & \resultcell{\underline{0.0730}}{} & \resultcell{\underline{0.0763}}{} & \resultcell{\underline{0.0937}}{} & \resultcell{\underline{0.1184}}{} \\
\cmidrule(l){2-8}
\rowcolor{paperblue!6}
 & \textbf{PHiT (ours)} & \resultcell{\textbf{0.0621}}{*} & \resultcell{\textbf{0.0685}}{*} & \resultcell{\textbf{0.0749}}{*} & \resultcell{\textbf{0.0780}}{*} & \resultcell{\textbf{0.0980}}{*} & \resultcell{\textbf{0.1233}}{*} \\
\midrule
\multirow{8}{*}{\rotatebox[origin=c]{90}{Beauty}} & Explicit CoT$_{\text{GREAM}}$ & \resultcell{0.0228}{} & \resultcell{0.0293}{} & \resultcell{0.0365}{} & \resultcell{0.0351}{} & \resultcell{0.0553}{} & \resultcell{\underline{0.0837}}{} \\
 & MiniOneRec & \resultcell{0.0232}{} & \resultcell{0.0295}{} & \resultcell{0.0358}{} & \resultcell{0.0352}{} & \resultcell{0.0542}{} & \resultcell{0.0795}{} \\
 & ReaRec & \resultcell{0.0201}{} & \resultcell{0.0255}{} & \resultcell{0.0307}{} & \resultcell{0.0296}{} & \resultcell{0.0464}{} & \resultcell{0.0673}{} \\
 & LC-Rec & \resultcell{0.0178}{} & \resultcell{0.0222}{} & \resultcell{0.0261}{} & \resultcell{0.0260}{} & \resultcell{0.0407}{} & \resultcell{0.0592}{} \\
 & SASRec & \resultcell{0.0159}{} & \resultcell{0.0195}{} & \resultcell{0.0229}{} & \resultcell{0.0232}{} & \resultcell{0.0343}{} & \resultcell{0.0480}{} \\
 & GRU4Rec & \resultcell{0.0144}{} & \resultcell{0.0190}{} & \resultcell{0.0242}{} & \resultcell{0.0226}{} & \resultcell{0.0370}{} & \resultcell{0.0573}{} \\
 & LASAR & \resultcell{\underline{0.0239}}{} & \resultcell{\underline{0.0303}}{} & \resultcell{\underline{0.0366}}{} & \resultcell{\underline{0.0365}}{} & \resultcell{\underline{0.0563}}{} & \resultcell{0.0813}{} \\
\cmidrule(l){2-8}
\rowcolor{paperblue!6}
 & \textbf{PHiT (ours)} & \resultcell{\textbf{0.0242}}{} & \resultcell{\textbf{0.0311}}{*} & \resultcell{\textbf{0.0386}}{*} & \resultcell{\textbf{0.0375}}{*} & \resultcell{\textbf{0.0591}}{*} & \resultcell{\textbf{0.0890}}{*} \\
\midrule
\multirow{8}{*}{\rotatebox[origin=c]{90}{Steam}} & Explicit CoT$_{\text{GREAM}}$ & \resultcell{0.0273}{} & \resultcell{0.0371}{} & \resultcell{0.0464}{} & \resultcell{0.0417}{} & \resultcell{0.0722}{} & \resultcell{0.1090}{} \\
 & MiniOneRec & \resultcell{0.0258}{} & \resultcell{0.0357}{} & \resultcell{0.0481}{} & \resultcell{0.0427}{} & \resultcell{0.0732}{} & \resultcell{0.1224}{} \\
 & ReaRec & \resultcell{0.0277}{} & \resultcell{0.0393}{} & \resultcell{0.0510}{} & \resultcell{0.0485}{} & \resultcell{0.0852}{} & \resultcell{0.1317}{} \\
 & LC-Rec & \resultcell{0.0253}{} & \resultcell{0.0384}{} & \resultcell{0.0509}{} & \resultcell{0.0432}{} & \resultcell{0.0727}{} & \resultcell{0.1126}{} \\
 & SASRec & \resultcell{0.0237}{} & \resultcell{0.0338}{} & \resultcell{0.0460}{} & \resultcell{0.0384}{} & \resultcell{0.0698}{} & \resultcell{0.1185}{} \\
 & GRU4Rec & \resultcell{0.0276}{} & \resultcell{0.0356}{} & \resultcell{0.0494}{} & \resultcell{0.0476}{} & \resultcell{0.0722}{} & \resultcell{0.1266}{} \\
 & LASAR & \resultcell{\underline{0.0326}}{} & \resultcell{\underline{0.0429}}{} & \resultcell{\underline{0.0548}}{} & \resultcell{\underline{0.0505}}{} & \resultcell{\underline{0.0862}}{} & \resultcell{\underline{0.1389}}{} \\
\cmidrule(l){2-8}
\rowcolor{paperblue!6}
 & \textbf{PHiT (ours)} & \resultcell{\textbf{0.0352}}{*} & \resultcell{\textbf{0.0470}}{*} & \resultcell{\textbf{0.0562}}{*} & \resultcell{\textbf{0.0561}}{*} & \resultcell{\textbf{0.0927}}{*} & \resultcell{\textbf{0.1414}}{*} \\
\bottomrule
\end{tabular*}
\par\smallskip
\parbox{\linewidth}{\footnotesize N denotes NDCG. Baseline values are retained from the LASAR manuscript. Gains use unrounded scores. PHiT results are averaged over five random seeds. $^{*}$ denotes a significant improvement over the strongest baseline for that dataset--metric ($p<0.1$).}
\end{table}

%% file: figures/alternatives/beauty_evidence.tex
\begin{table}[H]
\centering
\caption{Beauty component and supervision comparisons. MiniOneRec (SFT) provides the recommendation-only reference. Base adds eight thought tokens to recommendation-only SFT; each lower-block row independently adds the indicated supervision to Base. Bold marks the highest score. $^{*}$ denotes a significant PHiT improvement over Base under a two-sided paired $t$-test ($p<0.1$); the gain row reports PHiT improvements over MiniOneRec (SFT).}
\label{tab:t4}
\begin{minipage}[c]{.66\linewidth}
\centering
\small
\setlength{\tabcolsep}{2.4pt}
\renewcommand{\arraystretch}{1.13}
\begin{adjustbox}{max width=\linewidth}
\begin{tabular}{lccccc}
\toprule
 & \multicolumn{3}{c}{NDCG} & \multicolumn{2}{c}{HR} \\
\cmidrule(lr){2-4}\cmidrule(l){5-6}
Variant & @1 & @5 & @10 & @5 & @10 \\
\midrule
MiniOneRec (SFT) & .00903 & .02190 & .02860 & .03470 & .05540 \\
+ Thought tokens (Base) & .01050 & .02356 & .03051 & .03681 & .05848 \\
\midrule
Base + Shuffled targets & .01061 & .02375 & .03067 & .03698 & .05853 \\
Base + CoT alignment & .01037 & .02375 & .03056 & .03720 & .05849 \\
Base + CoT keywords & .01049 & .02373 & .03083 & .03687 & .05901 \\
\rowcolor{paperblue!6}
\textbf{PHiT} & \textbf{.01089}$^{*}$ & \textbf{.02416}$^{*}$ & \textbf{.03107}$^{*}$ & \textbf{.03751}$^{*}$ & \textbf{.05910}$^{*}$ \\
\midrule
Gain over MiniOneRec & $+20.6\%$ & $+10.3\%$ & $+8.6\%$ & $+8.1\%$ & $+6.7\%$ \\
\bottomrule
\end{tabular}
\end{adjustbox}
\end{minipage}\hfill
\begin{minipage}[c]{.32\linewidth}
\centering
\includegraphics[width=\linewidth]{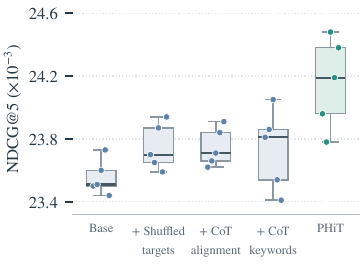}
\end{minipage}
\par\smallskip
\captionsetup{type=figure,hypcap=false}
\caption{\textbf{Five-seed NDCG@5 distributions on Beauty.} Each dot is one seed; boxes show the interquartile range and median, with whiskers extending to observations within 1.5 interquartile ranges. The vertical axis is truncated and scaled by $10^{-3}$. PHiT improves NDCG@5 over all four controls under two-sided paired $t$-tests ($p<0.1$). }
\label{fig:core_ablation}
\end{table}

%% file: figures/alternatives/cost.tex
\begin{table}[H]
\centering
\caption{Beauty supervision construction and Stage-2 training costs. NDCG@5 is averaged over five seeds.}
\label{tab:cost_main}
\small
\setlength{\tabcolsep}{8pt}
\renewcommand{\arraystretch}{1.12}
\begin{tabular*}{\linewidth}{@{\extracolsep{\fill}}lrr@{}}
\toprule
Supervision construction & Per target $\downarrow$ & Total (131,413 targets) $\downarrow$\\
\midrule
Qwen3.6-35B-A3B CoT generation & 66.20 ms & 145 min\\
\rowcolor{paperblue!6}
\textbf{PHiT} & \textbf{0.86 ms} & \textbf{113 s}\\
\bottomrule
\end{tabular*}
\smallskip
\begin{tabular*}{\linewidth}{@{\extracolsep{\fill}}lrr@{}}
\toprule
Method & Training time & NDCG@5\\
\midrule
Base & 2,467 s & .02356\\
+ CoT keywords & 2,741 s & .02373\\
\rowcolor{paperblue!6}
\textbf{PHiT} & \textbf{2,471 s} & \textbf{.02416}\\
\bottomrule
\end{tabular*}
\smallskip
\parbox{\linewidth}{\footnotesize PHiT and Qwen3.6-35B-A3B construction times are measured on the complete Beauty training set. Per-target costs amortize construction over the output targets. Stage-2 conditions use eight thought tokens and eight GPUs. Construction hardware and timing protocols are detailed in Appendix~\ref{app:cost_details}.}
\end{table}

%% file: sections/08_conclusion.tex
\section{Conclusion}
\label{sec:conclusion}

We introduced PHiT to learn thought tokens from observed item histories and execute them through standard causal prefill. A step-level diagnosis shows that history analysis is the only examined CoT component with positive mean contributions to both target-SID prediction and trace coherence; later preference and intent steps have negative mean answer contributions. PHiT uses cached item encodings to supervise contextual thought-token states, eliminating teacher-trace generation and recurrent hidden-state feedback. With SFT alone, it outperforms the compared baselines on four benchmarks. Controlled ablations support input-matched semantic supervision, and cost measurements show efficient target construction with Stage~2 training time comparable to Base.

%% file: sections/07_limitations.tex
\section{Limitations}
\label{sec:limitations}

PHiT is evaluated on offline next-item recommendation benchmarks built from item metadata and interaction histories. Future work can extend PHiT to multimodal item content, evolving catalogs, and larger-scale deployments. Our experiments instantiate PHiT on a Qwen3 backbone with hierarchical semantic identifiers; porting the history-informed thought tokens to additional generative backbones and identifier schemes would further broaden their applicability. We also plan to explore adaptive computation that adjusts the number of thought tokens to individual requests, balancing recommendation quality and inference cost.

%% file: sections/appendix.tex
\section{Experimental Settings and Data Processing}
\label{app:settings}

For the three Amazon benchmarks (Beauty, Instruments, and Sports), we directly adopt the preprocessed data splits released by GREAM~\citep{41_GREAM} as used in LASAR, without additional filtering or re-splitting; for Steam, we follow the exact dataset protocol of LASAR~\citep{lasar2026}. Across all main experiments, hyperparameters are tuned on the validation split and fixed prior to evaluating on the held-out test split, which reserves each user's final interaction for next-item prediction (dataset statistics in Table~\ref{tab:datasets}). Table~\ref{tab:configuration} summarizes the shared configuration of our five-seed controlled Beauty ablations.

\begin{table}[!htb]
\centering
\caption{Training configuration of the five-seed Beauty controls. All five conditions share the Stage~1 initialization and evaluation set.}
\label{tab:configuration}
\small
\begin{tabular*}{\linewidth}{@{\extracolsep{\fill}}ll@{}}
\toprule
Setting & Value \\
\midrule
Backbone & Qwen3-0.6B \\
Thought tokens & 8 fixed vocabulary IDs \\
Sparse encoder & \texttt{naver/splade\_v2\_max} (frozen) \\
Historical target size & 16 words per example \\
Historical prediction & Mean pooling; linear head; balanced BCE \\
Historical loss weight & 1.0 \\
Full-CoT control & Block-mean targets; MSE weight 0.1 \\
Beam width & 50 \\
Repeated-run seeds & 42, 43, 44, 45, 46 \\
Test examples & 22,363 \\
\bottomrule
\end{tabular*}
\end{table}

\paragraph{Historical target processing.}
To construct item representations, we concatenate each item's title, brand, categories, and description into a single text field and encode it with frozen \texttt{naver/splade\_v2\_max}~\citep{splade2021,spladev2_2021}. Subword WordPiece tokens are merged into normalized whole words, retaining only words that appear verbatim in the source item text. For each training instance, we sum the cached lexical weight vectors over the observed items in the user's history and select the top $K=16$ words to form the multi-hot target $z_u$. Targets use the observed history and exclude the held-out next item. In the \emph{+ Shuffled targets} control, we randomly permute the constructed target vectors $z_u$ across training examples via a derangement (preventing self-assignment), thereby preserving the exact marginal keyword distribution, prediction head, and loss function while severing the semantic link to the input history.

\paragraph{Recommendation metrics.}
Given a ranked list of predicted item SIDs where the ground-truth next item appears at 1-based rank $r$, Hit Rate at cutoff $k$ is defined as $\mathrm{HR}@k = \mathbf{1}[r\leq k]$, and Normalized Discounted Cumulative Gain at cutoff $k$ is defined as $\mathrm{NDCG}@k = \mathbf{1}[r\leq k]/\log_2(r+1)$ (both evaluate to zero when $r > k$). Reported metrics are averaged uniformly over all held-out test users. 

In Table~\ref{tab:t4}, MiniOneRec (SFT) is a single-run reference; the five thought-token conditions report five-run means.

\section{Target Construction, Training, and Inference}
\label{app:algorithm}

Algorithm~\ref{alg:pipeline} outlines the complete PHiT pipeline. Sparse item encodings are computed once and cached across all training histories, while test-time inference requires neither the sparse encoder nor the auxiliary keyword classification head.

\begin{algorithm}[htbp]
\caption{History supervision and standard-prefill recommendation}
\label{alg:pipeline}
\small
\begin{algorithmic}[1]
\Require Examples $\mathcal D$; item texts $\{d_i\}$; thought tokens $N$; keywords $K$
\State Encode and cache sparse word weights $a_i(w)$ for each item
\For{each $(H_u,y_u)\in\mathcal D$}
  \State Sum historical weights and retain top-$K$ observed words as $z_u$
\EndFor
\State Add $N$ distinct thought tokens to the tokenizer and model
\For{each training batch}
  \State Run a causal forward pass on history, fixed thought tokens, and target SIDs
  \State Predict $z_u$ from mean-pooled thought tokens; compute $\mathcal L_{\rm rec}+\lambda\mathcal L_{\rm hist}$
  \State Update the recommendation model and auxiliary head
\EndFor
\State Discard the auxiliary head for inference
\For{each inference request}
  \State Prefill history and fixed thought-token IDs; decode SIDs and map them to items
\EndFor
\end{algorithmic}
\end{algorithm}

\section{U1-Only Supervision on Instruments}
\label{app:u1_observation}

Figure~\ref{fig:u1_grouped} and Table~\ref{tab:u1_instruments} report the empirical comparison between full-CoT supervision and U1-only supervision on Instruments. Retaining only the history-analysis step reduces the mean supervision length from 2,377 characters (full CoT) to 921 characters ($U_1$-only), a reduction of over 61\%, while improving all six recommendation metrics.

\input{tables/u1_instruments}

\paragraph{Training and evaluation.}
The full-CoT and U1-only conditions use the LASAR framework with SFT alone and are initialized from the same KL Stage~1 backbone. LASAR in Table~\ref{tab:main_results} includes subsequent RL training. Stage~2 checkpoints are selected on the validation set and evaluated using the same test-time thought-token budget.

\section{A Case of Historical Semantic Supervision}
\label{app:historical_example}

Table~\ref{tab:historical_case} follows a Beauty example from history to the supervision target. The teacher trace mixes product evidence with preference and intent statements. CoT-derived keywords retain wording such as \emph{core}, \emph{pattern}, \emph{missing}, and \emph{loyal}; PHiT instead constructs its target directly from the observed item metadata, retaining terms such as \emph{shellac}, \emph{polish}, \emph{cnd}, and \emph{coat}. Figure~\ref{fig:motivation} uses a subset of this case's terms.

\begin{table}[!htb]
\centering
\caption{A Beauty case comparing teacher-derived and history-derived supervision. History titles are excerpted; the complete three-step teacher CoT and both keyword sets are shown. The reference next item is not used to construct the PHiT target.}
\label{tab:historical_case}
\small
\renewcommand{\arraystretch}{1.16}
\begin{tabular}{@{}p{.21\linewidth}p{.75\linewidth}@{}}
\toprule
Component & Example content\\
\midrule
Observed history\newline (title excerpts) & Shellac Remover Wraps; Now Foods Essential Oil, Lemon; Debra Lynn Professional Clear Pump Dispenser Bottle; CND Shellac Top and Base Set.\\
\midrule
Teacher $U_1$\newline (history analysis) & --- Strong focus on at-home gel/shellac manicures: UV gel light dryer, Shellac top/base coat set, remover wraps, manicure cleanup brush. \par --- Supporting accessories/tools for application and removal: pump dispenser bottle (for acetone/alcohol), cleanup brush for precision. \par --- Ancillary item (lemon essential oil) suggests DIY/home care, but core pattern centers on nail services. \par --- User appears to be assembling a complete Shellac system (curing, application, removal) but lacks color coats and some prep/finish liquids.\\
Teacher $U_2$\newline (preferences) & --- Prefers salon-quality, professional-grade nail products (CND Shellac system, pro UV lamp). \par --- Values efficient application and removal workflows (remover wraps, pump dispenser, precision brush). \par --- Likely brand-loyal to CND for gel/shellac. \par --- Practical, versatile choices that support repeated at-home manicures.\\
Teacher $U_3$\newline (intent) & --- Having the UV lamp and CND Shellac top/base implies the next functional need is the CND Shellac color coat(s) to complete the manicure system. \par --- Secondary needs could include pure acetone or isopropyl alcohol and lint-free wipes, but color is the primary missing component to perform full services.\\
\midrule
CoT keywords & gel, manicures, uv, dryer, shellac, remover, brush, precision, lemon, core, pattern, nail, missing, prep, cnd, loyal\\
\rowcolor{paperblue!6}
PHiT target & nail, beauty, shellac, polish, salon, gel, top, accessories, cnd, professional, remover, coat, uv, crisp, makeup, skin\\
\midrule
Next item\newline (reference) & CND Shellac Gel Nail Polish Zillionaire .25 Oz\\
\bottomrule
\end{tabular}
\end{table}

\section{Step-Level Diagnostic Results}
\label{app:diagnosis}

Table~\ref{tab:t1} reports mean and median step contributions on the Beauty training set. Both identify history analysis as the only step with positive answer and coherence contributions. Step deletion diagnoses predictive support; downstream retraining and shuffled-target controls assess the resulting supervision design.

\input{tables/table_01}

\section{Run-Level Consistency on Beauty}
\label{app:seed_consistency}

\begin{figure}[!htb]
\centering
\includegraphics[width=\linewidth]{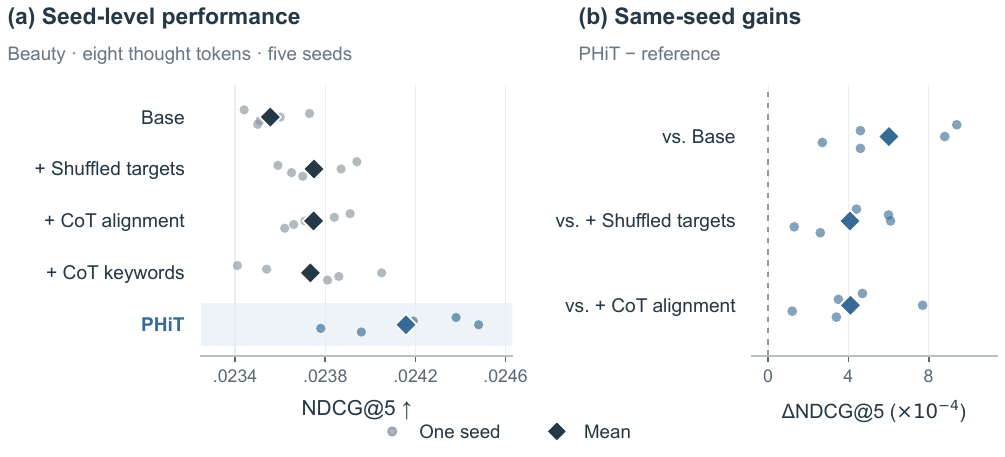}
\caption{\textbf{Run-level performance and same-seed gains on Beauty.} Eight thought tokens, five seeds. (a) Circles show individual NDCG@5 scores; diamonds show means. (b) Each circle is PHiT minus the indicated control for the same seed; diamonds show mean differences. The horizontal axis reports absolute NDCG@5 gains in units of $10^{-4}$: a value of 4 denotes 0.0004. The three displayed paired comparisons improve in all five runs; significance uses two-sided paired $t$-tests ($p<0.1$).}
\label{fig:paired_seed_details}
\end{figure}

Figure~\ref{fig:paired_seed_details} complements the main-text distributions with run-level scores and matched differences. PHiT has the highest mean NDCG@5 among the five conditions; its gain over Base, shuffled targets, and CoT alignment is positive in every matched run.

\section{Supervision and Training Cost Details}
\label{app:cost_details}

\begin{table}[H]
\centering
\caption{Beauty supervision construction and Stage~2 training times.}
\label{tab:supervision_summary}
\label{tab:t7}
\small
\setlength{\tabcolsep}{4pt}
\begin{adjustbox}{max width=\linewidth}
\begin{tabular}{@{}lllr@{}}
\toprule
Operation & Scale & Hardware & Time\\
\midrule
PHiT target construction & 131,413 targets & 1 GPU + CPU & \textbf{113 s}\\
\quad Item encoding & 12,101 items & 1 GPU & 77.8 s\\
\quad History aggregation & 176,139 histories & CPU & 30.0 s\\
Qwen3.6-35B-A3B CoT generation & 131,413 traces & 8 A800s & 145 min\\
\midrule
Base Stage~2 & Beauty training set & 8 GPUs & 2,467 s\\
+ CoT keywords Stage~2 & Beauty training set & 8 GPUs & 2,741 s\\
PHiT Stage~2 & Beauty training set & 8 GPUs & 2,471 s\\
\bottomrule
\end{tabular}
\end{adjustbox}
\par\smallskip
\parbox{\linewidth}{\footnotesize PHiT: three end-to-end runs. Qwen3.6-35B-A3B: 145 minutes for the full training set on eight A800s. Aggregation covers the processed history pool; PHiT outputs 131,413 training targets.}
\end{table}

%% file: tables/u1_instruments.tex
\begin{table}[!htb]
\centering
\caption{Historical supervision on Instruments. U1-only retains the history-analysis step, reducing supervision characters by over 61\%. Bold marks the best displayed scores. Run settings are in Appendix~\ref{app:u1_observation}.}
\label{tab:u1_instruments}
\small
\setlength{\tabcolsep}{5pt}
\begin{tabular}{lrrrrrr}
\toprule
 & \multicolumn{3}{c}{NDCG $\uparrow$} & \multicolumn{3}{c}{HR $\uparrow$} \\
\cmidrule(lr){2-4}\cmidrule(l){5-7}
Supervision & @5 & @10 & @20 & @5 & @10 & @20 \\
\midrule
Pure SFT & .0549 & .0609 & .0674 & .0710 & .0896 & .1153 \\
Full CoT & .0566 & .0620 & .0686 & .0718 & .0885 & .1147 \\
\rowcolor{paperblue!6}
U1-only & \textbf{.0569} & \textbf{.0632} & \textbf{.0699} & \textbf{.0727} & \textbf{.0922} & \textbf{.1191} \\
\bottomrule
\end{tabular}
\end{table}

%% file: tables/table_01.tex
\begin{table}[!htb]
\centering
\caption{Step-level utility on the Beauty training set. Medians are in parentheses. Coherence is undefined for the final step.}
\label{tab:t1}
\small
\setlength{\tabcolsep}{4pt}
\renewcommand{\arraystretch}{1.1}
\begin{adjustbox}{max width=\linewidth}
\begin{tabular}{lccc}
\toprule
Step & Function & $\Delta A$: mean (median) & $\Delta C$: mean (median) \\
\midrule
\textbf{1} & \textbf{History analysis} & \textbf{+0.0183 (+0.0006)} & \textbf{+0.0344 (+0.0657)} \\
2 & Preference construction & $-0.0627$ ($-0.0530$) & $-0.0766$ ($-0.0852$) \\
3 & Intent inference & $-0.0307$ ($-0.0806$) & N/A \\
\bottomrule
\end{tabular}
\end{adjustbox}
\end{table}